\documentclass[amssymb,amsfonts,aps,prl,twocolumn,floatfix]{revtex4-2}

\usepackage{graphicx}
\usepackage{dcolumn}
\usepackage{bm}
\usepackage[normalem]{ulem}
\usepackage{xcolor}
\usepackage{comment}
\usepackage{amsmath}
\usepackage{siunitx}

\DeclareSIUnit\angstrom{\text{Å}}

\begin{document}

\title{
Anatomy of linear and non-linear intermolecular exchange in $S=1$ nanographenes
}

\author{
 J. C. G. Henriques$^{1,2}$, 
 J. Fern\'andez-Rossier$^{1}$\footnote{On permanent leave from Departamento de F\'isica Aplicada, Universidad de Alicante, 03690 San Vicente del Raspeig, Spain.}
}

\affiliation{
(1) International Iberian Nanotechnology Laboratory (INL), Av. Mestre Jos\'e Veiga, 4715-330 Braga, Portugal
}
\affiliation{
(2) Universidade de Santiago de Compostela, 15782 Santiago de Compostela, Spain
}

\date{\today}

\begin{abstract} 
Nanographene triangulenes with a $S=1$ ground state have been  used as building blocks of antiferromagnetic Haldane spin chains realizing a symmetry protected topological phase. By means of inelastic electron spectroscopy, it was found that the intermolecular exchange contains both linear and non-linear interactions, realizing the bilinear-biquadratic Hamiltonian.  Starting from a Hubbard model, and mapping it to an interacting Creutz ladder, we analytically derive these effective spin-interactions using perturbation theory, up to fourth order. We find that for chains with more than two units  other interactions arise, with same order-of-magnitude strength, that entail second neighbor linear, and three-site non-linear exchange. Our analytical expressions compare well with experimental and numerical results. We discuss the extension to general $S=1$ molecules, and give numerical results for the strength of the non-linear exchange for several nanographenes. Our results pave the way towards  rational design of spin Hamiltonians for nanographene based spin chains.
\end{abstract}

\maketitle
Non-linear exchange, i.e., spin interactions that go beyond the simple Heisenberg coupling between two spins,  play a prominent role in many physical systems, such as antiferromagnetic transition metal oxides\cite{Anderson59}, magnetic impurities in insulators\cite{harris63},   magnetic multilayers\cite{slonczewski1991}, chiral magnets\cite{paul20} and magnetic two-dimensional materials\cite{kartsev2020}.  Non-linear exchange is a key ingredient in the exactly solvable AKLT models\cite{Affleck87}, whose ground state is a resource for measurement based quantum computing\cite{Wei2011}.

The relative size and sign of linear and non-linear exchange can have a dramatic impact in several contexts.
For instance,  whereas the swap gate, or permutation operator,  for spin qubits can be implemented with a linear Heisenberg interaction\cite{divincenzo2000}, for spin qudits  it requires the presence  of non-linear exchange terms\cite{segraves64}.
Alternatively, in the case of the two-dimensional $S=3/2$ honeycomb lattice, the relative size of linear and non-linear exchange controls the nature of the ground state and its excitation spectrum \cite{Affleck87,Pomata20,ganesh2011}.

Here we undertake the exploration of non-linear exchange in $S=1$ nanographenes. This class of system features  outstanding flexibility to realize $S=1$ molecules with different shapes and sizes. It has been recently shown\cite{yan2023} that there are 383 different nanographenes that can be formed with 9 hexagons or less. Therefore, this type of system provides an ideal arena to engineer intermolecular exchange. One of the simplest $S=1$ nanographenes is the so-called [3]-triangulene.

Triangulenes are graphene fragments with the shape of an equilateral triangle, of various sizes and terminated with zigzag edges; these are customarily defined in terms of the number of benzenes, $n$, in a given edge\cite{fernandez07,su2020} (termed a [n]-triangulene).  Based on Lieb's theorem for the Hubbard model for bipartite lattices at half-filling \cite{Lieb1989}, $[n]$-triangulenes are predicted to be open-shell multiradicals, with the spin of the ground state scaling as $2S=n-1$, associated with a half-full shell of $n-1$ in-gap non-bonding zero modes\cite{ovchinnikov78,fernandez07,wang2008,wang09,yazyev10,ortiz19}. 

Due to recent breakthroughs in bottom-up synthesis techniques \cite{ruffieux2016surface,mishra2019b,su2019,su2020triangulenes,mishra2021b}, and the capability of atomic precision manipulation of organic molecules, triangulenes have been used as building blocks of larger molecular structures \cite{mishra2020,Mishra2021,hieulle2021,cheng2022surface}. A prime example of this is the recent realization of Haldane spin chains, where [3]-triangulenes (henceforth referred simply as triangulene) were coupled in order to generate chains with more than 16 units \cite{Mishra2021}. There, the inelastic electron tunneling spectroscopy (IETS) was described with an effective $S=1$ spin Hamiltonian that included  both linear and non-linear exchange terms, i.e. the so called BLBQ Hamiltonian \cite{tanaka2018origin,hu2020,catarina2022,soni2022,Catarina23}:
\begin{align}
    H_\text{BLBQ} = \sum_{i} J \boldsymbol{S}_i \cdot \boldsymbol{S}_{i+1} + B \left( \boldsymbol{S}_i \cdot \boldsymbol{S}_{i+1} \right)^2,
    \label{eq: BLBQ}
\end{align} 
where each triangulene is represented by a spin-1 operator $\boldsymbol{S}_i$, the sum runs over all triangulenes in the chain and the parameters $J$ and $B$ are the linear and non-linear exchange couplings, respectively. 
The introduction of the non-linear exchange term proved essential to increase the accuracy of the spin model when compared with experimental data \cite{Mishra2021} and full fermionic numerical approaches \cite{catarina2022}. 

In the following, we derive the effective $S=1$ spin Hamiltonian   (\ref{eq: BLBQ}) starting from a fermionic model for the  nanographenes.  Importantly, our derivation unveils the presence of  additional second neighbor linear interactions, and   non-linear exchange that involve three-spin terms, with strength comparable to $B$ in Eq. (\ref{eq: BLBQ}). As a fermionic model  we  use a single-orbital  Hubbard model (see Supporting Information), containing hoppings between first ($t$) and third neighbor ($t_3$) sites, and an on-site Hubbard repulsion term ($U$) which deals with the intra-atomic Coulomb repulsion cost associated with having a given $\pi$-orbital doubly occupied \cite{ortiz19}. The Hubbard model has been  validated by comparison with multiconfigurational calculations 
obtained with full-quantum chemistry \emph{ab-initio} methods \cite{ortiz19}, for $t = -2.7$ eV, $t_3 \sim t/10$ and $U \sim |t|$ \cite{ortiz22}.

\begin{figure}
	\centering
	\includegraphics[width=\columnwidth]{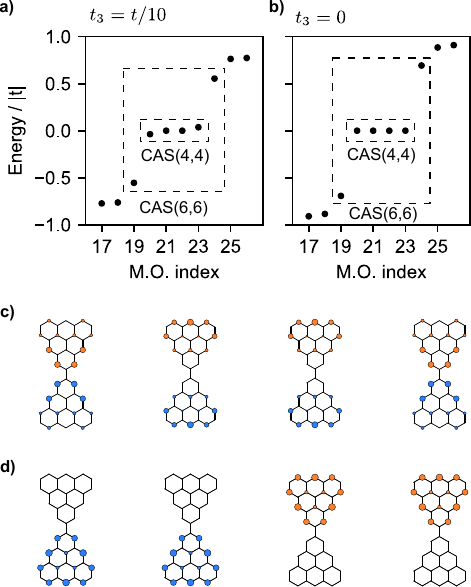}
	\caption{Single particle energies for the triangulene dimer with $t_3 = t/10$ (a) and $t_3 = 0$ (b). Energies are given in unit of $|t|=2.7$ eV. The dashed boxes indicate two possible choices for the active space where the Hubbard Hamiltonian is represented. CAS($N_O, N_e$) refers to an active space with $N_O$ modes where $N_e$ electrons are distributed). Panels (c) and (d) show the absolute value of the site representation of the four molecular orbitals closer to zero energy for $t_3 = t/10$ and $t_3 = 0$, respectively. Each circle is colored according to the sublattice where it is located.}
    \label{fig:Single_Particle}
\end{figure}

As a starting point, we employ a Hubbard Hamiltonian at half-filling to describe the smallest chain, i.e.
a triangulene dimer. In order to derive the effective spin model, we describe the fermionic many-body states  of the nanographenes in a truncated Hilbert space, corresponding to the complete active space (CAS) approximation (see Supporting Information) \cite{ortiz19,Mishra2021,jacob21,jacob22}. In Fig. \ref{fig:Single_Particle} (a) and (b) we show the single-particle energies for the triangulene dimer with $t_3 = t/10$ and $t_3 = 0$; in both cases two possible choices of active spaces are indicated. When $t_3 = 0$ one finds four states with zero energy; these correspond to the unhybridized zero modes of the individual triangulenes. For $t_3 = t/10$, intermolecular hybridization promoted by $t_3$ lifts this degeneracy. In panels (c) and (d) of the same figure, the absolute value of the site representation of the four modes closest to zero energy are depicted for $t_3 = t/10$ and $t_3 = 0$. For the latter case, the wave functions were chosen as eigenfunctions of the $C_3$ symmetry operator with eigenvalues $\pm 2\pi/3$. For future reference, we refer to panel (c) as the molecular orbital basis, and to panel (d) as the $C_3$ symmetric basis.

To gain physical insight about the properties of this system, we represent the Hamiltonian in the $C_3$ symmetric basis taking into account only the four modes at zero energy. The representation of the Hubbard model in that basis leads to the following effective Hamiltonian:
\begin{align}
    H &= \sum_{\mu' \mu} \sum_\sigma \tau_{\mu' \mu} d^\dagger_{\mu',\sigma} d_{\mu,\sigma} + \sum_\mu \mathcal{U_\mu} n_{\mu, \uparrow} n_{\mu, \downarrow}  \notag \\
    &+ \sum_\triangle \mathcal{J}_{\triangle_+, \triangle_-} \left( n_{\triangle_+,\uparrow} n_{\triangle_-,\downarrow} + n_{\triangle_-,\uparrow} n_{\triangle_+,\downarrow} \right) \notag \\
    &+ \sum_\triangle \mathcal{J}_{\triangle_+, \triangle_-} \left( d^\dagger_{\triangle_+,\uparrow} d_{\triangle_-,\uparrow} d^\dagger_{\triangle_-,\downarrow} d_{\triangle_+,\downarrow} + \text{h.c.} \right)
    \label{eq: CAS_Hamiltonian}
\end{align}
where $\mu$ and $\mu'$ run over the four $C_3$ symmetric modes and $\sigma = \uparrow, \downarrow$. The sums over $\triangle$ run over the two triangulenes, and $\triangle_\pm$ refers to the modes with eigenvalues $\pm 2\pi/3$ in a given triangulene. The operator $d^{(\dagger)}_{\mu,\sigma}$ annihilates (creates) an electron in the $\mu$ mode with spin $\sigma$, and $n_{\mu, \sigma} = d^{\dagger}_{\mu,\sigma} d_{\mu,\sigma}$. 

\begin{figure}
    \centering
    \includegraphics[width=\columnwidth]{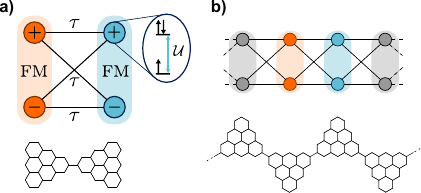}
    \caption{(a) Schematic representation of the interactions present in the Hamiltonian of Eq. (\ref{eq: CAS_Hamiltonian}), with modes in the same triangulene ferromagnetically coupled, and hoppings between the two triangulenes; (b) Extension of the model for the case of a triangulene chain.}
    \label{fig:Creutz Ladder + PT}
\end{figure}

This Hamiltonian contains three distinct types of terms. The first term in Eq. (\ref{eq: 
CAS_Hamiltonian}) represents the hopping between two modes, $\mu$ and $\mu'$, with an amplitude $\tau_{\mu\mu'} = t_3 \sum_{\langle\langle\langle i,j \rangle\rangle\rangle} \Phi^*_\mu(i) \Phi_{\mu'}(j)$ where $\Phi_\mu(i)$ is the site representation of the $\mu$-mode (depicted in Fig. \ref{fig:Single_Particle}). This hopping amplitude is zero when the two modes are localized in the same sublattice; thus, hoppings between modes in the same triangulene vanish. In addition, due to the $C_3$ symmetry of the $\mu$-modes, all the finite hoppings have the same absolute value, which we label simply as $\tau$. The pictorial representation of this hopping structure, shown in 
fig. \ref{fig:Creutz Ladder + PT}, reveals the single-particle Hamiltonian maps into a Creutz ladder model\cite{creutz1999end} with vanishing vertical hoppings. 

The remaining terms in the Hamiltonian account for electron-electron interactions. The second term of Eq. (\ref{eq: CAS_Hamiltonian}), corresponds to an effective Hubbard repulsion that penalizes the double occupancy of a given mode $\mu$, with the energy penalty being proportional to the inverse participation ratio (IPR), and is defined as $\mathcal{U}_\mu = U \sum_i |\Phi_\mu (i) |^4$, where the sum runs over all the sites. Again, due to the $C_3$ symmetry of the modes, one finds $\mathcal{U}_\mu \equiv \mathcal{U}$ is independent of $\mu$. 

The last two lines in Eq. (\ref{eq: CAS_Hamiltonian}) describe an intra-triangulene ferromagnetic exchange interaction, with the exchange coupling given by $\mathcal{J}_{\mu\mu'} = U \sum_i |\Phi_\mu (i) |^2 |\Phi_{\mu'} (i) |^2$. As before, $\mathcal{J}_{\mu\mu'} = \mathcal{J}$ is independent of $\mu$ and $\mu'$, and $\mathcal{U} = \mathcal{J}$. 

Therefore the  model (\ref{eq: CAS_Hamiltonian}) realizes a Creutz-ladder \cite{creutz1999end}, with the difference that the vertical hoppings are replaced by a ferromagnetic exchange and  there are Hubbard interactions.
In Fig. \ref{fig:Creutz Ladder + PT} we present a pictorial representation of the Hamiltonian of Eq. (\ref{eq: CAS_Hamiltonian}) for the triangulene dimer, as well as the extension to larger structures, e.g. a triangulene chain. The extension of Eq. (\ref{eq: CAS_Hamiltonian}) to triangulene chains is easily obtained by running the sums over all triangulenes (see Supporting Information for the numerical solution of Eq.(\ref{eq: CAS_Hamiltonian}) for a triangulene dimer and trimer). 

\begin{figure}
    \centering
    \includegraphics[width=\columnwidth]{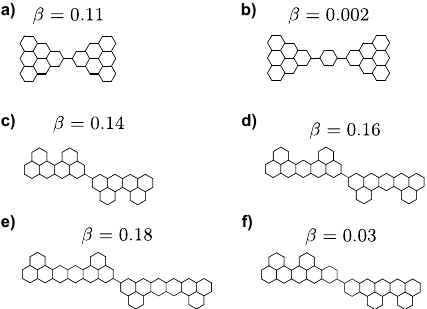}
    \caption{Values obtained for $\beta=B/J$ by numerically diagonalizing the Hubbard Hamiltonian in a restricted Hilbert space with 4 modes only, for dimers obtained from $S=1$ building blocks, taking $U=|t|$, $t_3=t/10$ and $|t|=2.7$eV.}
    \label{fig:beta various molecules}
\end{figure}

Let us note in passing, that for molecules less symmetric than triangulenes, for example panels (c) to (f) of Fig. \ref{fig:beta various molecules}, the hoppings $\tau_{\mu \mu'}$, the effective Hubbard repulsion $\mathcal{U}_\mu$, and the exchange coupling $\mathcal{J}_{\mu\mu'}$ are in principle mode-dependent. Moreover, it might be necessary to include a new term associated with electron-pair hoppings in Eq. (\ref{eq: CAS_Hamiltonian}). As discussed in \cite{ortiz19}, this term vanishes for triangulenes, but may be finite for other molecules. Including it is straightforward, and does not affect the main physical features we presently wish to discuss.

Starting from the many-body fermionic model just presented,  we now produce an effective low energy description which can be related to the effective BLBQ spin model. To this end, we shall employ degenerate perturbation up to fourth order, treating the hoppings $\tau$ as a perturbation. 

For the unperturbed system ($\tau=0$) the two pairs of modes are decoupled, and each triangulene can be studied individually. At half-filling, with two electrons per triangulene, the lowest energy states in a given triangulene are $|\uparrow\rangle|\uparrow\rangle$, $|\downarrow\rangle |\downarrow\rangle$ and $\left(|\uparrow\rangle |\downarrow\rangle + |\downarrow\rangle |\uparrow\rangle \right)/\sqrt{2}$, where each ket refers to one of the $C_3$ symmetric modes in a triangulene.
These states correspond to the three spin projections of a $S=1$ state formed by the ferromagnetic coupling of two spin-1/2 electrons, i.e. $|\uparrow\rangle|\uparrow\rangle \equiv |\Uparrow\rangle$, $|\downarrow\rangle|\downarrow\rangle \equiv |\Downarrow\rangle$ and $\left(|\uparrow\rangle |\downarrow\rangle + |\downarrow\rangle |\uparrow\rangle \right)/\sqrt{2} \equiv |+\rangle$, with the ket on the right hand side referring to the state of the triangulene as a whole. With these three states in each triangulene, the total unperturbed dimer Hamiltonian at half-filling is nine-fold degenerate.

Representing Eq. (\ref{eq: BLBQ}) in the basis of two spin-1 objects, and focusing on the $S_z = 0$ sector, one finds that while the bilinear term, proportional to $J$, is responsible for connecting the state $|+\rangle |+\rangle$ with the states $|\Uparrow\rangle |\Downarrow\rangle$ and $|\Downarrow\rangle |\Uparrow\rangle$, the biquadratic term, proportional to $B$, unlocks a new interaction between the states $|\Uparrow \rangle |\Downarrow\rangle$ and $|\Downarrow\rangle |\Uparrow \rangle$. Recalling the definition of the spin-1 states in terms of two spin-1/2, one realizes that while the processes mediated by $J$ link states which differ by two spin flips, the process mediated by $B$ connects states differing by four spin flips. Hence, the biquadratic interaction is only to be expected in 4th order perturbation theory, while the bilinear term should already be present in 2nd order.

The expressions for the 2nd and 4th order corrections in degenerate perturbation theory read \cite{calzado2004origin, malrieu2014magnetic}:
\begin{align}
    h^{(2)}_{kp} &= \sum_{m} \frac{\langle k | H_\tau | m \rangle \langle m | H_\tau | p \rangle}{\Delta E^{(0)}_m} \\
    h^{(4)}_{kp} &= \sum_{mm'm''} \frac{\langle k | H_\tau | m \rangle \langle m | H_\tau | m' \rangle \langle m' | H_\tau | m'' \rangle \langle m'' | H_\tau | p \rangle}{\Delta E^{(0)}_m \Delta E^{(0)}_{m'} \Delta E^{(0)}_{m''}} \notag \\
    & - \sum_{mm'l} \frac{\langle k | H_\tau | m \rangle \langle m | H_\tau | l \rangle \langle l | H_\tau | m' \rangle \langle m' | H_\tau | p \rangle}{\Delta E^{(0)}_m \Delta E^{(0)}_m \Delta E^{(0)}_{m'}}
\end{align}
where $k$, $p$ and $l$ label states inside the degenerate ground state, and the sums run over all the states outside that subspace. The perturbation is $H_\tau = \sum_{\mu\mu'}\sum_\sigma \tau_{\mu\mu'} d^\dagger_{\mu\sigma} d_{\mu\sigma}$ and $\Delta E^{(0)}_m$ is the unperturbed energy of the $|m\rangle$ state relative to the ground state. In the absence of applied magnetic field, the odd-order corrections vanish identically \cite{macdonald1988t}.
One important aspect to note is that even though the initial and final state correspond to open-shell configurations with two electrons per triangulene, the intermediate states contain closed-shell configurations, as well as charge excitations, with different number of electrons in the two triangulenes.

In 2nd order perturbation theory, where only two electron flips are considered, a finite contribution for $J$ is found, which for the present system simply reads $J^{(2)} = 2\tau^2/\mathcal{U}$ (see Supporting Information). This result is similar to the one usually found when when a Hubbard chain is mapped to a Heisenberg chain of antiferromagnetically coupled spins; the different numerical pre-factor steming from the different geometry of our system.

Progressing to 4th order perturbation theory, where processes involving up to four electron are considered, different paths appear which allow for $|\Uparrow \rangle |\Downarrow \rangle \rightarrow |\Downarrow\rangle  |\Uparrow \rangle$. Carrying out the necessary calculations, one finds that finite contributions appear for both $J$ and $B$. These are given by $J^{(4)} = 4\tau^4/\mathcal{U}^3$ and $B^{(4)} = 8\tau^4/\mathcal{U}^3$. The 2nd order contribution to $J$ dominates its 4th order counterpart in the physically relevant region of the parameter space, and $J^{(4)}$ may be neglected. Hence, combining the results from second and forth order perturbation theory, we find in leading order:
\begin{align}
    J \approx \frac{2\tau^2}{\mathcal{U}},\quad B \approx \frac{8\tau^4}{\mathcal{U}^3}, \quad \beta \equiv \frac{B}{J} \approx \frac{4\tau^2}{\mathcal{U}^2}.
    \label{eq: Analytical J and B}
\end{align}
Using the definitions of $\tau$ \cite{ortiz22} and $\mathcal{U}$ \cite{ortiz19}, and assuming $t_3\approx t/10$, one can estimate the strength of the quadratic exchange with respect to the linear one, characterized by $\beta$, using $\beta \approx 0.1 (t/U)^2$. This rough approximation yields a nice agreement with the value $\beta = 0.09$ that was found in the description of experimental data in \cite{Mishra2021}. This good agreement, together with the analytical expressions of Eq. (\ref{eq: Analytical J and B}), indicate  that non-linear exchange is a higher-order manifestation of the same underlying kinetic exchange mechanism\cite{Anderson59} that gives rise to linear exchange.

To validate our analytical expressions we shall now compare them with the results found from numerical diagonalization. The numerical results are obtained by first diagonalizing Eq. (\ref{eq: CAS_Hamiltonian}), followed by matching the energies of the first excitations with those of the BLBQ Hamiltonian (see Supporting Information for details). This gives numerically both $J$ and $B$ as a function of the parameters of the microscopic model, $U$ and $t_3$, and allows the comparison with Eq. (\ref{eq: Analytical J and B}).

\begin{figure}
    \centering
    \includegraphics[width=\columnwidth]{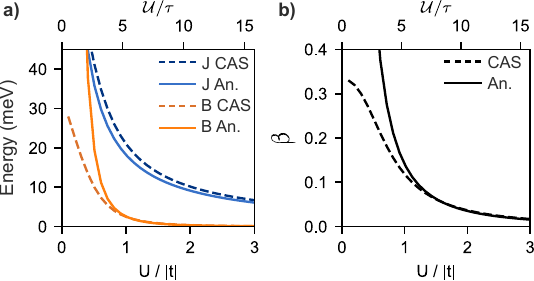}
    \caption{Comparison between the analytical expression and the result found from numerical diagonalization for (a) the linear exchange $J$ and the non-linear exchange $B$, and (b) the ratio $\beta=B/J$ as a function of the on-site Hubbard repulsion $U$ in units of $|t|= 2.7$ eV with $t_3 = t/10$.}
    \label{fig:Analytical_vs_CAS_Kinetic}
\end{figure}

In Fig. \ref{fig:Analytical_vs_CAS_Kinetic} we show the comparison between the analytical expressions found for $J$ and $B$ and the numerical results. 
Inspecting this figure reveals an excellent agreement between the two approaches, especially in the region $U\gg |t|$ where perturbation theory is valid. Crucially, this agreement holds near $U\sim |t|$, the physically relevant region. For $U=|t|$, one finds the linear exchange $J = 20$ meV and $\beta = 0.11$. As $U$ increases $\beta$ approaches 0, and a Heisenberg-like picture is recovered. For $U<<|t|$, outside the validity region of perturbation theory, the mapping to a spin model fails since the order of low energy excitations is no longer the same in the two approaches. The fact that $\beta$ is bounded by the AKLT limit $\beta=1/3$ is a consequence of Lieb's theorem, which prevents a four-fold degenerate ground state. 

Having completed the study of the triangulene dimer, we extend our analysis to larger chains.  Generalizing Eq. (\ref{eq: CAS_Hamiltonian}) to include more triangulenes, and once again employing perturbation theory up to 4th order, we find that for a chain composed of $N$ triangulenes the effective spin Hamiltonian reads (see Supporting Information for detailed derivation):
\begin{align}
    H_N &= J \sum_{i=1}^{N-1} \boldsymbol{S}_i \cdot \boldsymbol{S}_{i+1} + B \sum_{i=1}^{N-1} (\boldsymbol{S}_i \cdot \boldsymbol{S}_{i+1})^2 \notag \\
    & + J_2 \sum_{i=1}^{N-2} \boldsymbol{S}_i \cdot \boldsymbol{S}_{i+2} \notag \\
    & + B_{1,1} \sum_{i=1}^{N-2} (\boldsymbol{S}_i \cdot \boldsymbol{S}_{i+1})(\boldsymbol{S}_{i+1} \cdot \boldsymbol{S}_{i+2}) + \text{h.c.}.
\end{align}
While the first line corresponds to the BLBQ model for $N$ triangulenes, the second and third lines contain new exchange interactions; the former describes an antiferromagnetic second neighbor linear exchange, and the latter a ferromagnetic quadratic exchange involving two adjacent triangulene pairs. We note that the antiferromagnetic second neighbor linear exchange might promote frustration if its strength becomes comparable with $J$. If perturbation theory is extended up to sixth or eight order, additional exchange interactions appear; however, since these are in higher order of $\tau/\mathcal{U}$ we neglect their contribution. Crucially, the terms proportional to $J_2$ and $B_{1,1}$ appear in the same order of perturbation theory as $B$ and therefore must be accounted for. To leading order, we find:
\begin{align}
    J = \frac{2\tau^2}{\mathcal{U}}, B = \frac{8\tau^4}{\mathcal{U}^3}, J_2 = \frac{79\tau^4}{12\mathcal{U}^3}, B_{1,1}= -\frac{37\tau^4}{12\mathcal{U}^3}
\end{align}
Hence, a consistent description of triangulene spin chains should include this terms, missing in previous analysis\cite{Mishra2021}.

We now briefly explore the non-linear exchange of various $S=1$ dimers. By applying numerical diagonalization in the minimal CAS, and comparing it with the excitation energies of the BLBQ Hamiltonian, we obtain the value of $\beta$ for different dimers composed of molecules with a triplet ground state. This should predict the relevance of non-linear exchange, as well as its tunability, in chains formed with different building blocks. In Fig. \ref{fig:beta various molecules} (b) we show a triangulene dimer, where a benzene is introduced in the middle. Formally, this molecule is nearly identical to the dimer we considered in the text (depicted in panel (a)), with the main difference being that due to the extra benzene, intermolecular hybridization between the two triangulenes is greatly diminished, resulting in a smaller $\tau$. This leads to $\beta = 0.002$, two orders of magnitude smaller than what we found for the case without the spacer (see Supporting Information for details on the change in this value due to a larger active space). Next, we consider the molecule of panel (c), where each monomer is composed of two Phenalenyl side-by-side; for this dimer $\beta = 0.14$. As benzenes are added between the two Phenalenyl, depicted in panels (d) \cite{su2020atomically} and (e), the value of $\beta$ increases to $\beta = 0.16$ and $\beta = 0.18$, respectively. This increase in $\beta$ may be ascribed to the decrease of the IPR of the zero modes of the individual molecules. At last, we consider panel (f), where an additional benzene is added close to the binding site of the two molecules. This leads to a significant decrease of $\beta$ to $\beta = 0.03$. The reason for this sharp decrease is similar to the one we gave when discussing panel (b). Hence, we see that by carefully choosing the geometry of the molecules used as building blocks, it should be possible to engineer the strength of non-linear exchange.

Even though we restricted our analysis to the minimal CAS, accounting only for the zero modes, additional orbitals could have been included in the calculation (see Fig. \ref{fig:Single_Particle}). This would introduce the so-called Coulomb driven super-exchange discussed in \cite{jacob22}. We have verified that including this additional mechanism would slightly change the numerical results, but preserve the qualitative features of the model.

\emph{Conclusions:} We have considered chains of $S=1$ nanographenes, taking the case of triangulene chains as the prototypical example. Using perturbation theory we have found that non-linear exchange interactions are a higher-order manifestation of the same mechanisms that give rise to linear exchange, and we have obtained analytical expressions for their amplitude. The  analytical results permit us to  relate molecule geometry with $\beta$, the degree of exchange non-linearity.  In addition, our analysis with more than two molecules shows that new terms appear in the Hamiltonian, namely a second neighbor linear and a three-site non linear exchange interaction, going beyond the BLBQ paradigm. Future work will address the impact of these extra terms on the well established phase diagram of the BLBQ model.

We acknowledge fruitful discussions with Gonçalo Catarina, António Costa and David Jacob.
We  acknowledge financial support from 
 FCT (Grant No. PTDC/FIS-MAC/2045/2021),
 SNF Sinergia (Grant Pimag),
FEDER /Junta de Andaluc\'ia, 
(Grant No. P18-FR-4834), 
Generalitat Valenciana funding Prometeo2021/017
and MFA/2022/045,
and
 funding from
MICIIN-Spain (Grant No. PID2019-109539GB-C41).

\bibliographystyle{apsrev4-2}
\bibliography{main}

\end{document}


\maketitle


\section{Solving the Hubbard Hamiltonian for the triangulene dimer and trimer}

In this section we describe how to solve the Hubbard Hamiltonian in the configuration interaction approach within the complete active space (CAS) approximation and show results for both the triangulene dimer and trimer.

First, we write the Hubbard model as
\begin{align}
    H = t \sum_{\langle i,j\rangle} \sum_\sigma c^\dagger_{i,\sigma} c_{j,\sigma} + t_3 \sum_{\langle\langle\langle i,j \rangle\rangle\rangle} \sum_\sigma c^\dagger_{i,\sigma} c_{j,\sigma} + U \sum_i n_{i\uparrow} n_{i\downarrow}
\end{align}
where $t$ and $t_3$ refer to first and third neighbor hopping, respectively, $U$ is the on-site Hubbard repulsion parameter and $\sigma=\uparrow,\downarrow$. The operator $c^{(\dagger)}_{i,\sigma}$ annihilates (creates) an electron at site $i$ with spin $\sigma$, and $n_{i,\sigma} = c^\dagger_{i,\sigma} c_{i,\sigma}$.

Solving the Hubbard model exactly is only possible for small structures due to the exponential increase of the Hilbert space with the number of sites. To circumvent this problem, the configuration interaction approach combined with the CAS approximation is often employed \cite{ortiz19}. In this framework, the single particle problem is solved first. Then, the full Hamiltonian is expressed in terms of the single particle eigenstates (termed molecular orbitals), and $N_e$ electrons are distributed over a subset of $N_O$ orbital close to zero energy; the remaining electrons fully occupy the orbitals at lower energy (at charge neutrality, $N_e = N_O$).

\begin{figure}
    \centering
    \includegraphics[scale = 0.75]{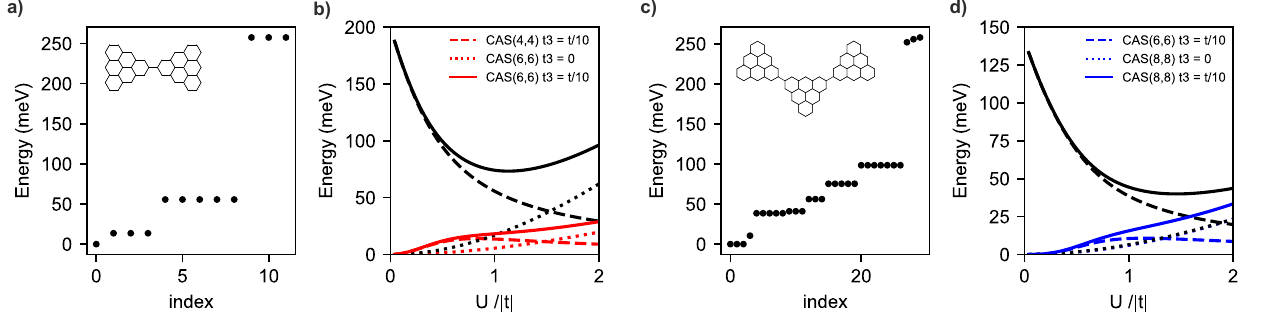}
    \caption{(a) Eigenvalues of the Hubbard Hamiltonian for the triangulene dimer (a) and trimer (c) obtained with CAS(4,4) and CAS(6,6), respectively, with $U=|t|$, $t_3 = t/10$ and $t=-2.7$eV. (b) Energy of the first two excited manifolds as a function of $U$ obtained with CAS(4,4) with $t_3 = t/10$ (dashed lines) and CAS(6,6) with $t_3 = 0$ (dotted lines) and $t_3 = t/10$ (solid lines). (d) Energy of the first two excited manifolds as a function of $U$ obtained with CAS(6,6) with $t_3 = t/10$ (dashed lines) and CAS(8,8) with $t_3 = 0$ (dotted lines) and $t_3 = t/10$ (solid lines).}
    \label{fig:CAS dimer}
\end{figure}

Let us now consider the case of a triangulene dimer. Its single particle spectrum is composed of 4 molecular orbitals at (or close to) zero energy, which are separated from the remaining ones by an energy gap of the order of $t$. Solving the Hubbard Hamiltonian at charge neutrality in an active space composed of these four zero modes only, i.e. CAS(4,4), we obtain the results depicted in Fig. \ref{fig:CAS dimer}a). In agreement with Lieb's theorem \cite{Lieb1989} and owing to the lack of sublattice imbalance of the dimer, the ground state is a singlet. The first two excited manifolds are a triplet, followed by a quintuplet. At higher energies, Well separated from these low energy excitations, other manifolds appear. 
In Fig. \ref{fig:CAS dimer}b) we depict the energy of the first two excited manifolds as a function of the Hubbard repulsion $U$ for $t_3 = t/10$. In the same figure we also show the results obtained with CAS(6,6), where the first occupied and unoccupied molecular orbitals were included in the active space, for $t_3 = 0$ and $t_3 = t/10$. Setting $t_3 = 0$ isolates the contribution of the so called Coulomb driven super exchange \cite{jacob22}. These results clearly show that including orbitals beyond the zero modes preserves the qualitative nature of the spectrum but introduces some quantitative changes especially for $U\gg |t|$; the region near $U\sim |t|$ is already fairly well described with the simpler CAS(4,4).

For a triangulene trimer, the single particle spectrum has 6 states close to zero energy. Solving the Hubbard model in the CAS(6,6) approximation, one finds a triplet ground state, followed by a set of excited manifolds at low energy, which are well separated from other more energetic excited states (see Fig. \ref{fig:CAS dimer} c).
As before, enlarging the active space preserves the qualitative features of the many-body solution, even though quantitatively some changes are noticeable (see Fig. \ref{fig:CAS dimer}d).

\section{Derivation of BLBQ Hamiltonian for triangulene dimers using perturbation theory}

In this section we describe in more detail how to obtain the BLBQ Hamiltonian, given in the main text, for a triangulene dimer using perturbation theory.

As discussed in the main text, the unperturbed Hamiltonian ($\tau=0$) for the triangulene dimer has a 9-fold degenerate ground state. Since the perturbation conserves $S_z$ we focus our analysis in a single $S_z$ sector, $S_z=0$. There are three states with $S_z = 0$ in the degenerate ground state manifold, namely $|\Uparrow \rangle |\Downarrow \rangle$, $|\Downarrow \rangle |\Uparrow \rangle$ and $|+ \rangle |+ \rangle$ where each ket refers to a triangulene, and $|\Uparrow \rangle = |\uparrow\rangle |\uparrow\rangle$, $|\Downarrow \rangle = |\downarrow\rangle |\downarrow\rangle$ and $|+ \rangle = (1/\sqrt{2})\left(|\uparrow\rangle |\downarrow\rangle + |\downarrow\rangle |\uparrow\rangle \right)$ where each ket on the right hand side refers to one of the two $C_3$ symmetric modes in a given triangulene. Using the equation given in the main text for the 2nd order correction to the Hamiltonian, and computing the matrix elements using the aforementioned three states with $S_z = 0$, one finds:
\begin{align}
    h^{(2)} = \frac{\tau^2}{\mathcal{U}}
            \begin{bmatrix}
                -4 & 2 & 0 \\
                2 & -2 & 2 \\
                0 & 2 & -4
            \end{bmatrix},
\end{align}
where the matrix is represented in the basis $|\Uparrow \rangle |\Downarrow \rangle$, $|+\rangle |+\rangle$, $|\Downarrow \rangle |\Uparrow \rangle$. At this point we note in passing that when performing this type of calculation one should be careful with the sign conventions in the definitions of the fermionic states. Representing the BLBQ Hamiltonian in the same basis, we find:
\begin{align}
    h_{\text{BLBQ}} = 
            \begin{bmatrix}
                -J + 2B & J-B & B \\
                J-B & 2B & J-B \\
                B & J-B & -J+2B
            \end{bmatrix}.
\end{align}
Demanding the two Hamiltonian to be equal (apart from some constant energy shift), we find the 2nd order contribution to $J$ and $B$ to be given by
\begin{align}
    J^{(2)} = 2\frac{\tau^2}{\mathcal{U}}, \quad B^{(2)} = 0,
\end{align}
as obtained in the main text.

In order to find the 4th order corrections to $J$ and $B$ a similar procedure is followed. Using the expression given in the main text for $h^{(4)}_{kp}$, we arrive at
\begin{align}
    h^{(4)} = \frac{\tau^4}{\mathcal{U}^3}
            \begin{bmatrix}
                0 & -4 & 8 \\
                -4 & 4 & -4 \\
                8 & -4 & 0
            \end{bmatrix}.
\end{align}
Once again, using the matrix representation of the BLBQ Hamiltonian as a reference, we obtain
\begin{align}
    J^{(4)} = 4\frac{\tau^4}{\mathcal{U}^3}, \quad B^{(4)} = 8\frac{\tau^4}{\mathcal{U}^3}.
\end{align}
As mentioned in the main text, because $J^{(4)}$ is of higher order in $\tau/\mathcal{U}$ than $J^{(2)}$, its contribution is rather small, and can be safely neglected. To better illustrate this let us consider some typical values for $\tau$ and $\mathcal{U}$. For triangulenes, one has $\tau=2t_3/11$ and $\mathcal{U}\approx0.1 U$. Taking $U=|t|$ and $t_3 = t/10$ as typical values, we find $J^{(4)}/J^{(2)} = 2 \tau^2/\mathcal{U}^2 \approx 0.06$, which makes $J^{(4)}$ more than one order of magnitude smaller than $J^{(2)}$.

\section{Extension to larger chains}
In this section we extend the results of the previous one to larger triangulene chains. We restrict our analysis to the chain with 3 triangulenes, a trimer, which already hosts the interactions which are absent in the dimer. 

For the triangulene trimer the ground state of the unperturbed system is a $3^3 = 27$ degenerate manifold. Of these, 7 states belong to the $S_z = 0$ subspace (check the number of low energy manifolds in Fig. \ref{fig:CAS dimer}c). These 7 states are the ones we will consider in the following analysis. Using the definition of $h^{(2)}$ given in the main text, and considering the basis $|\Uparrow \rangle |+ \rangle |\Downarrow \rangle$, $|\Uparrow \rangle |\Downarrow \rangle |+ \rangle$, $|+ \rangle |\Uparrow \rangle |\Downarrow \rangle$, $|+ \rangle |+ \rangle |+ \rangle$, $|+ \rangle |\Downarrow \rangle |\Uparrow \rangle$, $|\Downarrow \rangle |\Uparrow \rangle |+ \rangle$, $|\Downarrow \rangle |+ \rangle |\Uparrow \rangle$, we find
\begin{align}
    h^{(2)} = \frac{\tau^2}{\mathcal{U}}
            \begin{bmatrix}
                -4 & 2 & 2 & 0 & 0 & 0 & 0 \\
                2 & -6 & 0 & 2 & 0 & 0 & 0 \\
                2 & 0 & -6 & 2 & 0 & 0 & 0 \\
                0 & 2 & 2 & -4 & 2 & 2 & 0 \\
                0 & 0 & 0 & 2 & -6 & 0 & 2 \\
                0 & 0 & 0 & 2 & 0 & -6 & 2 \\
                0 & 0 & 0 & 0 & 2 & 2 & -4
            \end{bmatrix}.
\end{align}
In the same basis, the BLBQ Hamiltonian reads:
\begin{align}
    h_{\text{BLBQ}} = 
            \begin{bmatrix}
                2B & J & J & 0 & 0 & 0 & 0 \\
                J & 3B-J & 0 & J-B & 0 & B & 0 \\
                J & 0 & 3B-J & J-B & B & 0 & 0 \\
                0 & J-B & J-B & 4B & J-B & J-B & 0 \\
                0 & 0 & B & J-B & 3B-J & 0 & J \\
                0 & B & 0 & J-B & 0 & 3B-J & J \\
                0 & 0 & 0 & 0 & J & J & 2B
            \end{bmatrix}.
\end{align}
Once again demanding the two Hamiltonian to be equal apart from a constant shift in energy we find
\begin{align}
    J^{(2)} = 2\frac{\tau^2}{\mathcal{U}}, \quad B^{(2)} = 0.
\end{align}

Extending this analysis to 4th order, we find for $h^{(4)}$:
\begin{align}
    h^{(4)} = \frac{\tau^4}{12\mathcal{U}^3}
            \begin{bmatrix}
                -182 & 86 & 86 & 5 & 0 & 0 & 0 \\
                86 & -56 & -74 & -47 & 42 & 96 & 0 \\
                86 & -74 & -56 & -47 & 96 & 42 & 0 \\
                5 & -47 & -47 & 89 & -47 & -47 & 5 \\
                0 & 42 & 96 & -47 & -56 & -74 & 86 \\
                0 & 96 & 42 & -47 & -74 & -56 & 86 \\
                0 & 0 & 0 & 5 & 86 & 86 & -182
            \end{bmatrix}.
\end{align}
Comparing this matrix with $h_{\text{BLBQ}}$ it is clear that some entries which vanish in $h_{\text{BLBQ}}$ are finite in $h^{(4)}$; this clearly indicates that the BLBQ model is incomplete, and additional interactions should be accounted for in order to describe the triangulene trimer with an effective spin model. To determine which interactions are lacking, one should study which states yield a finite matrix element in $h^{(4)}$ which is not captured by the BLBQ Hamiltonian. For example, the states $|\Uparrow\rangle |+\rangle |\Downarrow\rangle$ and $|+\rangle |+\rangle |+\rangle$ give a finite matrix element in $h^{(4)}$. These two states are clearly connected via a second neighbor linear exchange which links the first and third triangulenes, i.e. $\boldsymbol{S}_1 \cdot \boldsymbol{S}_3$, and produces the spin flip $|\Uparrow\rangle |...\rangle |\Downarrow\rangle \rightarrow |+\rangle |...\rangle |+\rangle$. 

Applying the same type of reasoning to the other matrix elements that are not properly captured by the BLBQ model, we find that the appropriate spin Hamiltonian to describe the trimer up to 4th order in $\tau/\mathcal{U}$ reads:
\begin{align}
    H &= J (\boldsymbol{S}_1 \cdot \boldsymbol{S}_2 + \boldsymbol{S}_2 \cdot \boldsymbol{S}_3) 
       + B \left[ (\boldsymbol{S}_1 \cdot \boldsymbol{S}_2)^2 + (\boldsymbol{S}_2 \cdot \boldsymbol{S}_3)^2 \right] \notag \\
      &+ J_2 \boldsymbol{S}_1 \cdot \boldsymbol{S}_3 \notag \\
      &+ B_{1,1} \left[(\boldsymbol{S}_1 \cdot \boldsymbol{S}_2) (\boldsymbol{S}_2 \cdot \boldsymbol{S}_3) +  (\boldsymbol{S}_2 \cdot \boldsymbol{S}_3) (\boldsymbol{S}_1 \cdot \boldsymbol{S}_2) \right]
      \label{eq:Full spin Hamiltonian}
\end{align}
with the 4th order corrections to the coefficients being given by:
\begin{align}
    J^{(4)} = \frac{49}{12}\frac{\tau^4}{\mathcal{U}^3},\quad B^{(4)} = 8\frac{\tau^4}{\mathcal{U}^3},\quad J_2^{(4)} = \frac{79}{12}\frac{\tau^4}{\mathcal{U}^3}, \quad  B_{1,1}^{(4)} = -\frac{37}{12}\frac{\tau^4}{\mathcal{U}^3}.
\end{align}
Notice that while $B^{(4)}$ is the same that we found in the dimer, $J^{(4)}$ differs from the one we found before by $\tau^4/12\mathcal{U}^3$; we thus see that the presence of an additional triangulene slightly renormalizes the linear exchange between nearest neighbors. This is due to processes with four electron hoppings, where an electron visits the third triangulene, but leaves without changing its initial state. Although interesting, this renormalization is tiny, and we ignore it completely. We further note that this Hamiltonian contains an antiferromagnetic second neighbor interaction, and a ferromagnetic quadratic interaction involving two adjacent triangulene pairs.

If we applied the procedure we described so far to higher order corrections in perturbation theory, new interactions would, once again, appear in the Hamiltonian. An example of this is the second neighbor biquadratic exchange $(\boldsymbol{S}_1 \cdot \boldsymbol{S}_3)^2$. These new interactions, however, would produce minute changes in the effective description of the trimer, and as a result we drop them. 

In Fig. \ref{fig:fit to CAS}a we compare the energies obtained by fitting the eigenvalues of Eq. (\ref{eq:Full spin Hamiltonian}) to the many-body energies of the triangulene trimer obtained with CAS(6,6) (more details on this are given in the following section). The excellent agreement between the two data sets further justifies our decision of neglecting interactions which are of higher order in $\tau/\mathcal{U}$. In panels (b) and (c) of Fig. \ref{fig:fit to CAS}  we compare the values obtained for the parameters of the spin model with the fitting procedure to the analytical expressions found from perturbation theory. The agreement between the two approaches is excellent, both in magnitude and sign, especially for $U>|t|$.

The extension to even larger chains can be obtained in a straightforward manner using the same procedure. Doing so, one finds that for a triangulene chain with $N$ triangulenes, the effective spin description, up to 4th order in $\tau/\mathcal{U}$, reads
\begin{align}
    H_N &= J \sum_{i=1}^{N-1} \boldsymbol{S}_i \cdot \boldsymbol{S}_{i+1} + B \sum_{i=1}^{N-1} (\boldsymbol{S}_i \cdot \boldsymbol{S}_{i+1})^2 \notag \\
    & + J_2 \sum_{i=1}^{N-2} \boldsymbol{S}_i \cdot \boldsymbol{S}_{i+2} \notag \\
    & + B_{1,1} \sum_{i=1}^{N-2} (\boldsymbol{S}_i \cdot \boldsymbol{S}_{i+1})(\boldsymbol{S}_{i+1} \cdot \boldsymbol{S}_{i+2}) + (\boldsymbol{S}_{i+1} \cdot \boldsymbol{S}_{i+2})(\boldsymbol{S}_i \cdot \boldsymbol{S}_{i+1}),
\end{align}
which trivially follows from the Hamiltonian we found for the trimer by increasing the number of triangulenes.

\section{Comparison between spin model and Hubbard model solved with CAS}

In this section we briefly discuss how we numerically validate the analytical results found with perturbation theory. 

For the case of the triangulene dimer we found that the appropriate spin model was the BLBQ Hamiltonian. Solving this Hamiltonian analytically one finds a singlet ground state, followed by a triplet with energy $J - 3B$ and a quintuplet with energy $3J - 3B$ relative to the ground state. Since the number of variables in the spin model is the same as the number of low energy excitations, a linear system can be established which relates $J$ and $B$ with the CAS energies. Solving this system for several values of $U$ and $t_3$ one numerically obtains the dependence of $J$ and $B$ with these parameters, which can then be compared with the analytical expressions found with perturbation theory (as showed in the main text).

\begin{figure}
    \centering
    \includegraphics[scale = 0.75]{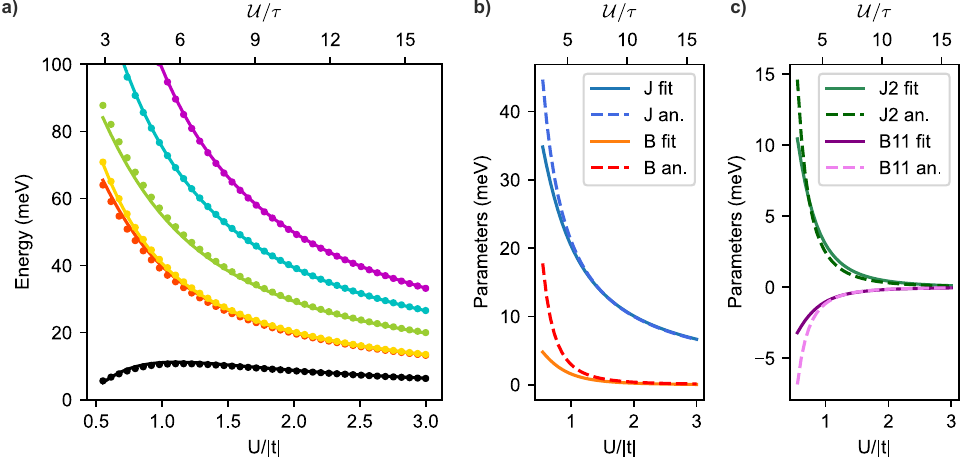}
    \caption{(a) Comparison between the first six excitation energies obtained by solving the Hubbard model for the triangulene trimer with CAS(6,6) with $t_3 = t/10$ (dots) and the ones obtained by fitting the spin model to the CAS results (lines); (b) and (c) Comparison between the parameters of the spin model obtained from fitting to CAS with the analytical expressions obtained with perturbation theory.}
    \label{fig:fit to CAS}
\end{figure}

For the triangulene trimer, we found that the model Hamiltonian has four parameters, $J$, $B$, $J_2$ and $B_{1,1}$. Contrarily to the case of the dimer, the low energy spectrum of the trimer has more excitation energies than there are parameters in the spin model (this is a consequence of truncating the Hamiltonian at 4th order). Thus, it is not possible to write a linear system of equations which relates the parameters of the spin model with the CAS excitation energies. Instead, we obtain the dependence of the parameters with $U$ and $t_3$ by fitting the eigenenergies of the spin model to those of CAS, taking the parameters of the spin model as fitting parameters. The parameters so obtained can then be compared with the analytical expressions (testing their validity). In Fig. \ref{fig:fit to CAS}a we compare the energies obtained by fitting the spin model to the CAS calculation for different values of $U$ and fixed $t_3$.
The agreement between the two approaches is clear, and it validates the spin model proposed by us. Also depicted, in panels (b) and (c), is the comparison between $J$, $B$, $J_2$ and $B_{1,1}$ obtained from the fitting procedure and the analytical ones. Once again, the two data sets show an excellent agreement.

\section{Comparison between spin and Hubbard Hamiltonians for a 4-site model}

\begin{figure}
    \centering
    \includegraphics[scale = 0.7]{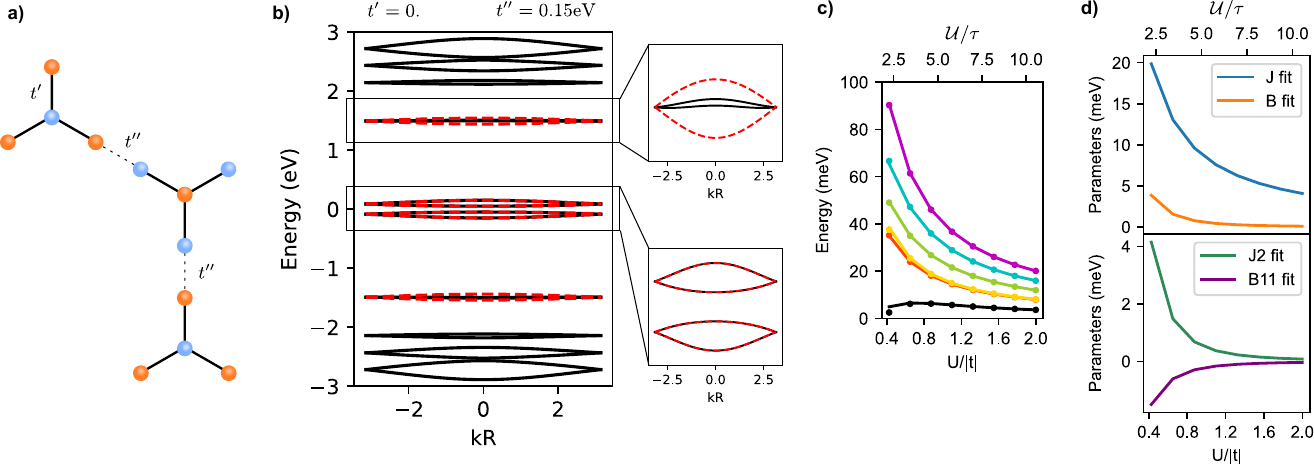}
    \caption{(a) Schematic representation of a four-site model trimer. (b) Bands obtained for an infinite triangulene chain (solid black lines) with $t=-2.7$eV and $t_3=t/10$, and for a chain of 4-site models (red dashed lines) obtained with $t'= -0.86$eV and $t''=-0.15$eV.(c) Comparison between the excitation energies obtained with exact diagonalization of the Hubbard model and the energies obtained from the spin Hamiltonian fit. (d) Parameters of the spin model obtained from the fitting procedure.}
    \label{fig:toy model}
\end{figure}

In this section we benchmark the effective spin Hamiltonian against the Hubbard model for a 4-site model trimer (depicted in Fig. \ref{fig:toy model}a) instead of the complete triangulene trimer.
The 4-site model (i.e. the building block) has been successfully used in the past to emulate triangulenes, since its sublattice imbalance and $C_3$ symmetry endows this system with main properties of a [3]triangulene \cite{catarina2022,Mishra2021}. The fundamental advantage of this simplified model is that, due to its reduced number of sites, exact diagonalization can be used. This should allow us to rule out the possibility that the spin interactions present in the model we proposed are spurious artifacts stemming from the constraints imposed by the active space.

The values of $t'$ and $t''$ are determined by computing the bands of an infinite chain \cite{ortiz22}, and fitting them to the bands of a triangulene chain with $t=-2.7$eV and $t_3=t/10$. In Fig. \ref{fig:toy model}b we depict the comparison between the bands of the two chains, obtained for $t'=-0.86$eV and $t''=-0.15$eV. 

With the parameters of the 4-site model fixed, we compute the energies for the trimer via exact diagonalization. Then, we fit the spin model to these results as described in the previous section. In Fig. \ref{fig:toy model}c we show the fit of the energies to the ones obtained with exact diagonalization; an excellent agreement is observed. In panel d) of the same figure we show the parameters of the spin model obtained from the fitting procedure; the signs and relative magnitudes are similar to those found for the triangulene trimer. These observations confirm that the spin interactions in the spin model are not artifacts arising from a restriction in the Hilbert space when the triangulene trimer was studied.

\section{The effect of increasing the active space on the value of $\beta$}

\begin{figure}
    \centering
    \includegraphics{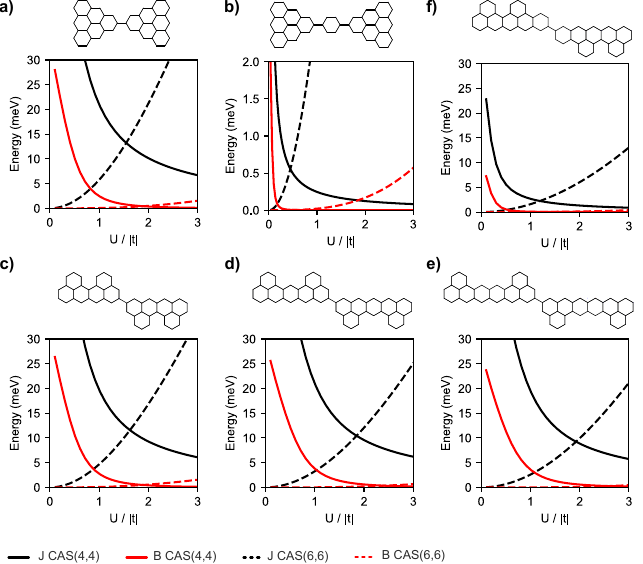}
    \caption{Values for $J$ (black) and $B$ (red) as a function of $U$ obtained by comparing the BLBQ spectrum with CAS calculations for the six dimers shown in the main text. Solid lines represent results obtained from CAS(4,4) with $t_3 = t/10$ and dashed lines correspond to results obtained with CAS(6,6) with $t_3 = 0$.}
    \label{fig:S=1_J_and_B}
\end{figure}

At last, in this section, we perform a more detailed study on the values of $\beta$ for the molecules discussed in the main text. We shall compute both $J$ and $B$ for several values of $U$ in two different active spaces, i.e. CAS(4,4) which accounts only for the 4 modes closest to zero energy, and CAS(6,6) which includes the highest occupied molecular orbital and the lowest unoccupied molecular orbital. In the CAS(4,4) calculation we use $t_3 = t/10$, while in the CAS(6,6) calculation we consider $t_3 = 0$. In the relevant region of the parameter space, i.e. $U \sim |t|$ the different exchange mechanisms are additive, and the results for CAS(6,6) with $t_3 = t/10$ are essentially given by summing the results from CAS(4,4) with $t_3=t/10$ with the ones from CAS(6,6) with $t_3 = 0$ \cite{jacob22}. These results are summarized in Fig. \ref{fig:S=1_J_and_B}; in Table \ref{tab:beta_values} we show the corresponding value of $\beta$ obtained for $U=|t|$ with the two considered active spaces with $t_3 = t/10$.
\begin{table}[]
\begin{tabular}{@{}lllllll@{}}
\toprule
                 & a & b & c & d & e & f \\ \midrule
$\beta_\text{CAS(4,4)}$ & $0.11$  & $0.002$  & $0.14$  & $0.16$ & $0.18$ & $0.03$  \\
$\beta_\text{CAS(6,6)}$ & $0.11$  & $0.01$  & $0.14$ & $0.16$ & $0.16$ & $0.03$  \\ \bottomrule
\end{tabular}
\caption{Values of $\beta=B/J$ obtained by comparing the BLBQ energies with CAS(4,4) and CAS(6,6) calculations, with $U=|t|$ and $t_3 = t/10$, for the six molecules mentioned in the main text.}
\label{tab:beta_values}
\end{table}

From Table \ref{tab:beta_values} we see that the values of $\beta$ obtained with CAS(4,4) and CAS(6,6) are similar in all cases, except for the triangulene dimer separated with a benzene, i.e. case (b). For this particular molecule, the benzene that separates the two triangulenes leads to a substantial reduction of intermolecular hybridization, as can be seen in Fig. \ref{fig:S=1_J_and_B} where the values of $J$ and $B$ are one rder of magnitude smaller than the ones of case (a) where the benzene spacer is absent. Due to the very weak intermolecular hybridization of the zero modes, the contribution from CAS(4,4) is not enough to accurately describe the system in the region $U \sim |t|$, and processes involving orbitals higher/lower in energy must be accounted for, as evidenced by the difference between CAS(4,4) and CAS(6,6) calculations.

Regarding the remaining molecules depicted in Fig. \ref{fig:S=1_J_and_B} we note that the contribution to exchange coming from CAS(4,4) is the most relevant one for $U \sim |t|$. However, as $U$ increases, the contribution from Coulomb driven exchange becomes more relevant and can even surpass the one coming from the hybridization of zero modes.

\bibliographystyle{apsrev4-2}
\bibliography{main}